\documentclass[11pt]{article}
\usepackage{graphicx,amscd,amsmath,amssymb,verbatim}

\usepackage{algpseudocode}   
\usepackage{algorithm} 
\usepackage{textcomp} 

\usepackage{tabularx} 
\usepackage{multirow}  
\usepackage{amsthm} 

\usepackage{authblk} 

\usepackage[margin=1.0in]{geometry} 

\newlength{\knuthian} 
\def\mathpalsave#1{\let\wasmathstyle=#1\relax}
\def\utilde#1{\mathpalette\mathpalsave
              {\setbox124=\hbox{$\wasmathstyle#1$}%
\setbox125=\hbox{$\fam=3\global\knuthian=\fontdimen5\font$}
\setbox125=\hbox{$\widetilde{\vrule height 0pt depth 0pt width \wd124}$}%
               \baselineskip=1pt\relax
               \vtop{\copy124\copy125\vskip -\knuthian}}}
\def\uhat#1{\mathpalette\mathpalsave
              {\setbox124=\hbox{$\wasmathstyle#1$}%
\setbox125=\hbox{$\fam=3\global\knuthian=\fontdimen5\font$}
\setbox125=\hbox{$\widehat{\vrule height 0pt depth 0pt width \wd124}$}%
               \baselineskip=1pt\relax
               \vtop{\copy124\copy125\vskip -\knuthian}}}      
\def\ubar#1{\mathpalette\mathpalsave
              {\setbox124=\hbox{$\wasmathstyle#1$}%
\setbox125=\hbox{$\fam=3\global\knuthian=\fontdimen5\font$}
\setbox125=\hbox{$\overline{\vrule height 0pt depth 0pt width \wd124}$}%
               \baselineskip=1pt\relax
               \vtop{\copy124\copy125\vskip -\knuthian}}}

\begin{document}


\title{\bf{A Hierarchical Exact Accelerated Stochastic Simulation Algorithm}}
\author[1]{David Orendorff}
\author[1,2]{Eric Mjolsness\thanks{emj@uci.edu}}
\affil[1]{Department of Computer Science, University of California, Irvine, USA}
\affil[2]{Institute for Genomics and Bioinformatics and Department of Mathematics}

     \maketitle

\begin{abstract}

A new algorithm, ``HiER-leap'', is derived 
which improves on the computational properties
of the ER-leap algorithm
for exact accelerated simulation of stochastic chemical kinetics. 
Unlike ER-leap, HiER-leap utilizes a hierarchical or divide-and-conquer
organization of reaction channels 
into tightly coupled ``blocks'' and is thereby able to speed up systems with many reaction channels. 
Like ER-leap, HiER-leap is based on the use of upper and lower bounds on the 
reaction propensities to define a rejection sampling algorithm with
inexpensive early rejection and acceptance steps.
But in HiER-leap,
large portions of  
intra-block sampling may be done in parallel. An accept/reject
step is used to synchronize across blocks. 
This method scales well when many reaction channels are present and has 
desirable asymptotic properties. The algorithm is exact,
parallelizable and achieves a significant speedup over SSA and ER-leap
on certain problems.
This algorithm offers a potentially important step towards efficient 
{\it in silico} modeling of entire organisms.
\end{abstract}

\begin{center}
\noindent Paper accepted for the Journal of Chemical Physics, 
published as \\
 http://link.aip.org/link/?JCP/137/214104 , 
DOI 10.1063/1.4766353.
\end{center}

\section{Introduction}

Computational biology is moving toward ever more complex, comprehensive and
detailed biological models. It is becoming increasingly important to simulate and understand these
models computationally. 
The Stochastic Simulation Algorithm\cite{Gillespie} (SSA) was introduced to exactly sample the
Chemical Master Equation and has seen widespread adoption. 
 
The original SSA iteratively samples reaction events in a way that requires 
$O(R)$ computational steps per sampled reaction event, 
where $R$ is the number of reaction channels. 
This can be prohibitively slow when there are a large number of reaction channels or reaction events.

This fact together with the importance of the SSA has inspired a slew of SSA acceleration techniques
\cite{GibsonBruck,Cao_Gillespie_Petzold_2005,Lu20121427,Gillespie_Petzold_2003,Jenkins20112580,Cangelosi:2010:SEL:1808143.1808190,Bayati:2009:DAS:1563054.1563219,Zhou_Peng_Yan_Wang_2008,Mjolsness_Orendorff_Chatelain_Koumoutsakos_2009}. 
The work of Gillespie\cite{tau-leap} and its recent variants \cite{ChatterjeeBinomialTauLeap2005,CGPtau05,CGPtau06}
reduces the total number of reaction events that need to be sampled but does so at the cost of accuracy. 
Additionally, the work of Gibson and Bruck \cite{GibsonBruck} reduces the amount of work per simulated reaction event to $\log R$. 
The work of Slepoy et al.\cite{Slepoy} ups the ante further by finding the next reaction event to sample in $O(1)$ time 
using rejection sampling under assumptions reasonable for biochemical networks. 

There have been recent advances in consumer level multi-core CPU technology. 
There are indications that next-generation CPU technology is moving
from maximizing single-core speed to increasing the number of cores by orders of magnitude. 
There has been work on the parallelization of SSA via GPUs
\cite{petzold-gpu,Jenkins20112580,Klingbeil_Erban_Giles_Maini_2011} and
multicore CPUs \cite{Gabrieli_Demontis_Pazzona_Suffritti_2011}.
However, the parallelization was used to speed up
sampling of {\it many} trajectories rather than speeding up each
trajectory in a large system.  
Multicore GPUs and CPUs have not been effectively used to speed up the
sampling of a single Chemical Master Equation trajectory exactly. Arguably this becomes the dominant problem
when extremely large
systems are being studied. For example, the {\it E. coli} genome has been
estimated to have about 4400 gene products\cite{Serres01012004}. 
This fact 
 suggests that tens of thousands of molecular species will needed
to be present if an {\it E. coli} specimen is to ever be comprehensively modeled {\it in
silico}.

Relatively little work has succeeded in reducing the number of reaction events 
sampled 
without introducing bias.
While the work of Riedel and Bruck\cite{RiedelBruck} is able to skip over cyclic states (eg loops), this method of reducing 
work does not apply to reaction networks with little state cycling. 
The previous work of the present authors (ER-leap) \cite{Mjolsness_Orendorff_Chatelain_Koumoutsakos_2009} is a leaping algorithm and 
was the first known general method 
to effectively reduce the number of SSA iterations sampled without sacrificing accuracy. 
This method scales well when reducing the number of SSA iterations. 
However, this method does not scale well when many reaction channels are present. 

The currently proposed work describes a new SSA-equivalent algorithm that can take advantage of
parallel hardware, 
and {\it additionally} provides an algorithmic speedup for systems with many
reaction channels. Like ER-leap, this ``HiER-leap'' (Hierarchical Exact Reaction-Leaping) 
algorithm achieves these advances without the loss of accuracy. 
The HiER-leap algorithm uses a divide-and-conquer strategy to independently sample 
sparsely connected submodules of the reaction network, in a way somewhat similar to ER-leap.
HiER-leap then performs a network-wide synchronization using rejection sampling. 
As will be shown, this synchronization step is efficient for "reasonably" independent submodules. The acceptance probability associated with
synchronization is asymptotically equal to one as the number of reaction channels goes to infinity.
This implies that the majority of the work will take place during submodule sampling, 
which may be performed in parallel. 

This work therefore presents a potentially important step towards organism-scale
simulation.

\section{Background Theory}
\label{background_theory}

The new HiER-leap algorithm begins its derivation from the state transition distribution
defined by the Chemical Master Equation after $L>0$ reaction events.
We then algebraically manipulate the CME until a distribution suitable for parallel 
sampling and synchronization is found.

In many ways this derivation closely follows the derivation found in ER-leap. 
Therefore, this section is dedicated to recalling the notation and key 
equations from ER-leap \cite{Mjolsness_Orendorff_Chatelain_Koumoutsakos_2009} that will serve as a starting point for the
algorithm derivation in section \ref{theory}.

\subsection{Notation}
\label{erleap_notation}

We define reaction channels, indexed by $r$, as a set of input and output species, $C_a$, with corresponding input
($m^{(r_a)}$) and output (${m'}^{(r_a)}$) stoichiometries
\begin{equation}
\left\{ m_{a}^{(r)} C_{a}\right\}  \longrightarrow  \left\{  {m^{\prime
}}_{a}^{(r)} C_{a}\right\} \ \ \ \text{\boldmath $\mathbf{with}$}\ \mathrm{reaction} \ 
\mathrm{rate} \  \rho _{r}
\label{XRef-Equation-stoch}
\end{equation}
and the net stoichiometry for a given species and reaction channel as
\[
\Delta m^{(r_a)}={m'}^{(r_a)}-m^{(r_a)}.
\]
Later, we will show the probabilities of state transitions after $L$ ``reaction events'' occur.

Under the Chemical Master Equation it is assumed that each reaction channel 
has a small probability of firing during a 
small time interval $dt$ with probability 
equal to $a_r(\mathbf{n})dt$. 
The vector $\mathbf{n}$, possibly indexed by $\alpha$ for species type $\alpha$, represents the 
quantities of the constituent species
in terms of
raw counts.
This $a_r(\mathbf{n})$ term is also called the {\it propensity} or 
{\it rate} of reaction channel $r$ and is defined as
\begin{gather}
a_{r}( \text{\boldmath $n$}) \equiv \rho _{r}F_{\text{\boldmath
$n$}}^{(r)} ,\notag \\
a_{0}( \text{\boldmath $n$}) \equiv \sum \limits_{r=1}^{R}a_{r}(
\text{\boldmath $n$}) 
\label{XRef-Equation-103112038}
\end{gather}
where 
\begin{equation}
\label{F_eq}
F_{\mathbf{n}}^{(r)}\equiv \prod \limits_{\left\{
a|m_{a}^{(r)}\neq 0\right\} }\left( \begin{cases}
\frac{n_{a} !}{\left( n_{a} -m_{a}^{(r)}\right) !} & \mathrm{if}\ n_{a}
\geqslant m_{a}^{(r)} \\
0 & \mathrm{otherwise} \\
\end{cases}\right).
\end{equation}
 
Note that superscripts involving ``$r$'' and related variables that index reaction numbers occur here and numerous times in the following. These
are enclosed in parentheses  ``$(r)$'' throughout, to indicate they are not powers but rather indexes.
 
In this work, it will be notationally convenient for us to keep
 the propensity term factored out into $\rho_r$ and $F^{(r)}_{\mathbf{n}}$.
 
As a brief aside, the upcoming derivations in this paper {\it may} work with other forms for $F^{(r)}_{\mathbf{n}}$. 
For example, the ``umbral transformation'' of a Hill function, 
\begin{align*}
F^{(r)}_{\mathbf{n}}&=Umbral[Hill(n;K)] \\
&= \frac{n_{(k)}}{(K^n + n_{(k)})},
\end{align*}
may work as propensity function, where the falling factorial $n_{(k)} \equiv \frac{n!}{(n-k)!}$ replaces $n^k$ in $Hill(n;K)$ 
(or more generally in a rational function) for each power of any integer-valued molecule number $n$. 
This functional form has the advantage of being monotonic and equal to zero for $n<k$, as required for a 
stochastic version of the Hill function with discrete integer numbers of molecules.
  
Furthermore, we define the total propensity $D_I$ for some reaction to occur in state $I$ as
\begin{equation}
\label{D_compute_eq}
D_I \equiv \sum_{r}\rho_r F^{(r)}_I ,
\end{equation}
which is equivalent to equation (\ref{XRef-Equation-103112038}).
 
Bounds for $F^{(r)}$ and $D$, after $L$ reaction events, are computed by bounding 
species counts after $L$ reaction events. 
For each species identifier (ID) $a$, we bound the number of molecules present after these reaction events, $n_{a}^{\prime}$, by:
\begin{equation}
n_{a}+L\ {\min }_{r}\left\{ \Delta  m_{a}^{(r)}\right\} \leqslant n_{a}^{\prime}\leqslant n_{a}+L\ {\max }_{r}\left\{ \Delta  m_{a}^{(r)}\right\}.
\label{n_tilde_bounds}
\end{equation}
If we introduce the notation that a tilde superscript or subscript, $\tilde{x}$ or $\utilde{x}$, represents upper
or lower bounding values respectively then, 
we can re-write the above as:
\begin{equation*}
\utilde{n}_a \leq n_{a}^{\prime} \leq \tilde{n}.
\end{equation*}
The corresponding propensities calculated from using the upper and lower bounding 
states, after $L-1$ reaction events, respectively are written as
\begin{align*}
\tilde{F}^{(r)}_{K,L-1} &\equiv F^{(r)}_{\tilde{\mathbf{n}}}\\
\utilde{F}^{(r)}_{K,L-1} &\equiv F^{(r)}_{\utilde{\mathbf{n}}}
\end{align*} 
and therefore
\begin{align*}
\tilde{D}_{K, L-1}=\sum_r \rho_r \tilde{F}^{(r)}_{K,L-1} \\
\utilde{D}_{K, L-1}=\sum_r \rho_r \utilde{F}^{(r)}_{K,L-1} 
\end{align*}
for any state $K$.

\subsection{Markov Process}
\label{erleap_markov_process}

In the ER-leap paper\cite{Mjolsness_Orendorff_Chatelain_Koumoutsakos_2009} it was shown that the probability of starting at state $I_0$ and ending up
in state $I_L$ after $\tau$ time elapses is
\begin{equation}
\label{L_event_prob}
P(I_L,\tau|I_0, L) = \sum \limits_{\left\{ R_{k}|k=1 .. L-1\right\} }
\left[ \prod \limits_{k=L-1\searrow 0}
	\rho_{R_k}{F_{I_{k}(\mathbf{R},I_{0})}^{\left( R_{k} \right) }}
	\exp ( -\tau_{k}( D_{I_{k}( \mathbf{R},I_{0}) ,I_{k}( \mathbf{R},I_{0}) }
) )  \right] 
\end{equation}
for every ordered vector $\{R_k|k=1\ldots L-1\}$ of reaction channel events. 
Each state may be uniquely transformed by a reaction 
as $I_0 \rightarrow I_1(R,I_0)$.

Furthermore, it was shown in ER-leap \cite{Mjolsness_Orendorff_Chatelain_Koumoutsakos_2009} that for any function $e(\mathbf{r})$ summing over all
possible orderings of $L$ reaction events is equivalent to summing over all possible counts of reactions (eg a multinomial with $L$ draws)
and then permuting each of these draws for all unequal reactions, yielding
\begin{equation}
\label{permutation_eq}
\sum \limits_{\left\{ r_{k}|k=1 .. L-1\right\} }e( \mathbf{r}) =
\sum \limits_{\left\{ \mathbf{s} | s_{r}\in \mathbb{N}, \sum _{r}s_{r}=L\right\} } 
\sum \limits_{\left\{\sigma  | \sigma \  \mathrm{permutes}\  \mathrm{unequal}\  r\mathrm{\mbox{'}}\mathrm{s} \mathrm{|} \text{\boldmath $s$}\right\} }e( \sigma( \mathbf{r}) )
\end{equation}
which, when combined with equation (\ref{L_event_prob}), and
introducing the  previously defined bounds, separating out terms in $e(\mathbf{r})$ which are permutation 
invariant, and after some algebra results in 
\begin{multline}
\label{erleap_update_eq}
P(I_L,\tau|I_0, L)=
\sum \limits_{\left\{
\text{\boldmath $s$} | s_{r}\in \mathbb{N} , \sum _{r}s_{r}=L\right\}}
\binom{L}{s_{1}\ \ \ ...\ \ \ s_{R}} 
\times \left[ \prod \limits_{r=1}^{R}{\left(
   \frac{\rho _{r}{\tilde{F}}_{I_0, L}^{\left( r\right) }}{\sum
\limits_{r}\rho _{r}{\tilde{F}}_{I_0, L}^{\left( r\right) }}
\right) }^{s_{r}}\right] \\
\times
\left(\utilde{D}_{I_{0} L-1}\right)^L
\exp \left( -\left(  \sum \limits_{k}\tau _{k} \right){\utilde{D}}_{I_{0} L-1} \right) 
\frac{{\left( {\tilde{D}}_{I_{0}L-1}\right) }^{L}}{\left(\utilde{D}_{I_{0} L-1}\right)^L} \\
\times {\left\langle  \left[ \prod \limits_{k=L-1\searrow 0}\left(
\frac{F_{I_{k}( \sigma ( \text{\boldmath $r$}\text{\boldmath $)$}
,I_{0}) }^{\left( r_{k}\right) }}{{\tilde{F}}_{I_{0},L-1}^{\left(
r_{k}\right) }}\right) \exp ( -\tau _{k}( D_{I_{k}( \sigma ( \text{\boldmath
$r$}\text{\boldmath $)$} ,I_{0}) ,I_{k}( \sigma ( \text{\boldmath
$r$}\text{\boldmath $)$} ,I_{0}) }-{\utilde{D}}_{I_{0}
L-1}))  \right] \right\rangle  }_{\left\{ \sigma |\text{\boldmath
$s$}\right\} } .
\end{multline}
This expression can be interpreted as a rejection-sampling
algorithm (last line) that corrects a multinomial 
approximate sampling algorithm (first two lines).

\subsubsection{Rejection Sampling}
Through the lens of rejection sampling, equation (\ref{erleap_update_eq}) represents an algorithm.

Briefly, rejection sampling is a method to sample $x$ from some distribution, $x \sim P(x)$,  
by means of an approximate distribution $P'(x)$. 
This can be expressed algebraically since $P(x)$ can be rewritten as 
\begin{equation}
\label{rejectionSamplingEq}
P( x) =P^{\prime }( x) \frac{P( x) }{M P^{\prime }( x) }+\left(
1-1/M\right) P( x) 
\end{equation}
assuming $M \geq 1$. 
Equation (\ref{rejectionSamplingEq}) can be viewed as a mixture distribution 
with the probability of sampling from $P'(x)$
being the ``acceptance'' $A(x)$:
\begin{equation}
A(x)=\frac{P(x)}{MP'(x)} 
\end{equation}
for some constant $M$ such that $\forall_x A(x)\leq 1$.
 
It is now possible to see the ER-leap algorithm represented in equation (\ref{erleap_update_eq}).
If in equation (\ref{erleap_update_eq}) we recognize $P(x)=P^{\prime }( x) M A( x)$ (equivalent to equation (\ref{rejectionSamplingEq})), with 
$M=\left( {\tilde{D}}_{I_{0}L-1}\right)^L / \left(\utilde{D}_{I_{0} L-1}\right)^L$, 
then we will implicitly define a $P'(x)$.
This $P'(x)$ has the next $L$ reaction events sampled from a multinomial
with the probability $p_r$ of choosing the $r^{th}$ reaction channel being equal to
\begin{equation*}
p_r=\left(
   \frac{\rho _{r}{\tilde{F}}_{I_0, L}^{\left( r\right) }}{\sum
\limits_{r}\rho _{r}{\tilde{F}}_{I_0, L}^{\left( r\right) }}
\right).
\end{equation*}
Furthermore, our $P'(x)$ samples $\tau$ from an Erlang distribution (equivalent to a Gamma distribution with integer ``shape'' parameter)
with rate parameter being $\utilde{D}_{I_0, L}$ and shape parameter being $L$.
Finally, if needed when calculating $A(x)$, a random permutation $\sigma$ is drawn uniformly and
$\{\tau_k|\tau=\sum_{k=0}^{L-1}\tau_k\}$ is sampled from an $L$-simplex.

The work in section \ref{theory} will similarly arrive at an equation 
representing an efficient and exact leaping algorithm for sampling $L$ reaction
events from an SSA equivalent distribution.

\section{Theory}
\label{theory}

\subsection {Hierarchical Notation}
\label{hierarchical-notation}

The HiER-leap algorithm uses a divide-and-conquer strategy 
to accelerate SSA.
Evidence suggests that protein-protein interaction (PPI) networks 
tend to be modular \cite{Maslov03052002}. These networks contain submodule clusters that
interact heavily inside the cluster. Interactions with other clusters of proteins are less common. 
Although still an active area of research, evidence \cite{10.1371/journal.pcbi.0040023} suggests that
similar modularity may exist in genetic regulatory networks as well. 
Additionally, when modeling spatial interactions \cite{Marquez-Lago2007, Rossinelli08a, Vlachos2008, Iyengar2010, citeulike:8603030,
citeulike:9639391}, events spatially distant must interact through sparse intermediate diffusion reaction channels.
In this way, it is probably common that many reaction channels are weakly coupled to the majority of other
channels. 
This observation suggests a potential avenue towards algorithm acceleration and
parallelization for large biological networks. 

Notation is introduced below to describe a hierarchical organization of reaction
channels. 
Table (\ref{table-notation}) provides a comprehensive guide to notation used throughout the following sections.
Next, following and generalizing the strategy of section \ref{background_theory},
we will derive bounds on propensities and
species.  The bounds will be essential for deriving an algorithm for exact
speedup of SSA for systems amenable to hierarchical organization. 

Reaction channels must belong to exactly one {\it block}. 
A {\it block} is defined as a set of reaction channels. 
If reactions are ``connected'' by shared reactants, it is 
preferred that reactions should be more strongly connected within
than between blocks.
For this work, a two
level hierarchy of reactions and blocks is used. However, it is straightforward to 
apply this method repeatedly to multiple levels. 

Each reaction channel is indexed by its block ID $r_1$, and its within-block ID
$r_2$, and will be designated as $R=(r_1 r_2)$ for $r_1 \in \{1\ldots b \}$ and $r_2 \in \{1 \ldots b_{r_1} \}$.  
The ``block propensity'' for block $r_1$ and state $I$, denoted $D^{(r_1)}_I$ is the sum of 
propensities of constituent reaction channels. Specifically, 
similar to equation (\ref{D_compute_eq})
this means
\begin{equation}
\label{D_block_eq}
D^{(r_1)}_I = \sum_{r_2 \in r_1} \rho_{r_1 r_2}F^{\left( r_1 r_2\right)}_I.
\end{equation}

Furthermore, we denote the number of reaction events occuring within block $r_1$ as
$u_{r_1}$. Finally, the number of events for the reaction channel indexed by
$R=(r_1 r_2)$ is denoted by $v_{r_1 r_2}$.

\subsection{Bounds on Propensities and Species Counts}
\label{dtilde-hierleap}

Similar to equation (\ref{n_tilde_bounds}), we now develop bounds on species counts
and propensities. This enables us to derive a two-scale rejection sampling algorithm in many
ways analogous to ER-leap at each scale.
For reasons that will become evident in section \ref{hierleap_derivation}, we first 
derive bounds on the block propensities given $L$ and $I_0$. Afterwards, bounds 
will be developed on the
species molecule counts and reaction channel propensities given $u$. 

First, recall that in equation (\ref{n_tilde_bounds}) 
we found bounds on species and propensities after $L$ reaction events.
Note that similar to equation (\ref{n_tilde_bounds}) we can define 
\begin{equation*}
\tilde{D}^{(r_1)}_{K,L-1} = \sum_{r_2 \in r_1} \rho_{r_1 r_2} \tilde{F}^{\left( r_1 r_2\right)}_{K,L-1}
\end{equation*}
and a similar definition for $\utilde{D}^{(r_1)}_{K,L-1}$.

\subsubsection{Optimized Block Level Bounds}

If it is the case that we only need bounds on the block propensities, and not
individual reaction channels, then we can take advantage of ``reaction event exclusion''.
This means that we only need to consider the sequence of at most length $L$ reaction events
which will result in the most extreme value for the {\it sum} of propensities in block $r_1$.
Therefore, we no longer need to assume that all species counts are at the
most extreme value possible after $L$ reaction events.

We want to find a bound closer to the optimal block propensity
\begin{equation}
{{}\widehat{D}^{(r_1)}_{(I_0 L)}}^*= \max_{v_{r_1}||u_{r_1}=L} 
	\sum_{r_2 \in r_1}{\rho_{r_1 r_2} F^{(r_1 r_2)}_{I(I_0, \mathbf{v}_{r_1})}}. 
\end{equation}
Unfortunately, na\"{\i}vely solving this exactly for $r_1$ requires
enumerating ${(b_{r_1})}^L$ possible choices for $\mathbf{v}_{r_1}$ upon every iteration.
Fortunately the bound we seek, ${}\widehat{D}^{(r_1)}_{(I_0 L)}$, is not required to be
exactly optimal. Instead we only require that
\begin{equation}
{{}\widehat{D}^{(r_1)}_{(I_0 L)}}^{*} \leq \widehat{D}^{(r_1)}_{(I_0 L)} \leq {\tilde{D}^{(r_1)}_{(I_0 L)}}
\end{equation}
such that ${{}\widehat{D}^{(r_1)}_{(I_0 L)}}^{*} \leq \widehat{D}^{(r_1)}_{(I_0 L)}$ is required for 
algorithmic correctness and 
$\widehat{D}^{(r_1)}_{(I_0 L)} \leq {\tilde{D}^{(r_1)}_{(I_0 L)}}$ is needed for improved efficiency.

A heuristic algorithm for $\widehat{D}^{(r_1)}_{(I_0 L)}$ is developed. 
We demonstrate this falls between the requisite values and has `nice' asymptotic
properties that will be discussed later. 

\begin{table}
ÊÊÊÊ\begin{tabular}{c l} \hline 
    	 \textbf{Symbol} & \textbf{Meaning} \\ \hline
		\multirow{2}{*}{$\tilde{x}$} & Upper bounding value for $x$ after $L-1$ reaction events. \\ 
			& Calculated by assuming each species type will be maximal. \\ \hline
		\multirow{3}{*}{$\hat{x}$} & Upper bounding value for $x$ after $L-1$ reaction events. \\
			& May not depend on bounding all species values and \\
			& therefore may be tighter than $\tilde{x}$. \\ \hline
    	\multirow{3}{*}{$\bar{x}$} & Upper bounding value given $u$. \\
    		& Will often involve inner block calculations. \\ \hline
    	Ê$\utilde{x}, \uhat{x}, \ubar{x} $ & Lower bounding versions of the above definitions.ÊÊÊÊÊ \\ \hline
    	$x^*$ & The optimal value of $x$ with respect to some objective function.  \\ \hline 
ÊÊÊÊ\end{tabular}
	\caption{Notation: Accents and Meaning}
	\label{table-notation}
\end{table}

\paragraph{Derivation}

The idea is to find the maximum $\Delta \widehat{D}^{(r_1)}_{(I_0 L)}$ possible
resulting from one reaction channel firing sometime during the next $L$ reaction events.
If we determine this value, we can upper bound $D^{(r_1)}_{(I_0 \ldots I_{L-1})}$ with
\begin{equation} 
\label{near_optimal_block_bound}
D^{(r_1)}_{(I_0,L)} \leq 
D^{(r_1)}_{I_0} + (L-1){\Delta \widehat{D}^{(r_1)}_{(I_0 L)}}^*
\end{equation} 
where   
\begin{equation*}
{\Delta \widehat{D}^{(r_1)}_{(I_0 L)}}^* = 
\max_{R_{r_1 r_2||I_0..I_{L-1}}} \Delta \widehat{D}^{(r_1)}_{(I_0 L)}. 
\end{equation*}
Note how this is an upper bound on $D^{(r_1)}_{(I_0,L)}$. By
construction, 
${\Delta \widehat{D}^{(r_1)}_{(I_0 L)}}^*$ is the largest amount that the block
propensity may change for any of the upcoming possible $(L-1)$ reaction events in $r_1$. 
Since there are $(L-1)$ reaction events, and the most any of them may
increase $D^{(r_1)}_{(I_0,L)}$ is 
${\Delta \widehat{ D}^{(r_1)}_{(I_0 L)}}^*$,   
equation (\ref{near_optimal_block_bound}) will always bound $D^{(r_1)}_{(I_0,L)}$.

This method improves upon our previous methods, which found the maximum $\tilde{n}_a$ 
for all species and then calculates the block propensity.
Each within-block reaction channel propensity will be larger when $\tilde{n}_a$ 
rather than $n_a$ is used to calculate the propensity. Therefore, using the increased bound
will result in a block's propensity being $O(b_{r_1} * L)$ larger than $D^{(r_1)}_{I_0}$. 
However, by calculating using  equation (\ref{near_optimal_block_bound}) 
the bound will be just $O(L)$ larger than $D^{(r_1)}_{I_0}$.    

Again, na\"{\i}vely solving for 
${\Delta \widehat{D}^{(r_1)}_{(I_0 L)}}^*$ 
requires an impractical amount of work. 
But as with our previous argument, we can upper-bound  
 ${\Delta \widehat{D}^{(r_1)}_{(I_0 L)}}^*$
 and still achieve an upper bound for $D^{(r_1)}_{(I_0,L)}$. 
 To upper bound 
 ${\Delta  \widehat{D}^{(r_1)}_{(I_0 L)}}^*$
 we use the monotonic nature of $D^{(r_1)}_{I_k}$. 
 If any species increases to $n_a' \geq n_a$ we know that 
 $D^{(r_1)}_{n_a'} \geq D^{(r_1)}_{n_a}$.
 Therefore, if we find the reaction channel that increases the block propensity
 the most when $\tilde{n}_{I_0, L}$ is used for positive $\Delta m^{(r_1r_2)}_a$, we are guaranteed that there 
 does not exist a larger $\Delta \widehat{D}^{(r_1)}_{(I_0 L)}$.
 
 This yields  
 \begin{equation}
 \label{delta_used_for_alg}
 \Delta \widehat{D}^{(r_1)}_{(I_0 L)} = \max_{r_2 \in r_1} \left[ D^{(r_1)}(\mathbf{q}(\mathbf{\tilde{n}}, r_2)) - D^{(r_1)}(\tilde{\mathbf{n}})
 \right]
 \end{equation}
 where 
 \begin{equation}
  \label{delta_used_for_alg2}
 \mathbf{q}(\mathbf{n}, r_2)_a=
\begin{cases} n_a+\Delta m^{(r_1r_2)}_a & \text{if $\Delta m^{(r_1r_2)}_a > 0$,} \\
n_a & \text{otherwise}
\end{cases}
 \end{equation}
 as our final equation for $\Delta \widehat{D}^{(r_1)}_{(I_0 L)}$.
 A proof that this bounds the maximum delta possible can be found in the appendix. 
 
This tighter bound will result in a greater acceptance ratio.
The basic reason for this improvement is that we need not overestimate 
{\it every} propensity in $r_1$ by $O(L)$, and add the overestimates up,
since only $L$ and not $b_1L$ reactions will occur. 
 
 Na\"{\i}vely finding the reaction channel $R={(r_1 r_2)}$ that will increase 
 $D^{(r_1)}(\tilde{n}_{I_0, L})$ by a maximal amount will
 cost $O(|R_{r_1}|)$ steps to compute.
 To accelerate this step further, blockwise priority queues (PQ) are used to find this $R$ efficiently.
 Nodes in the PQ are reaction channels and values are the $\Delta D^{(r_1)}_{(I(\tilde{n},  b_{r_1}))}$
 caused by each reaction channel firing. 
 Upon acceptance of $L$ reaction events we must update the priority queue for each block. 
 Only nodes that interact with species which have changed, need to be adjusted. 
 This, at worst, will be $O(\log{b_{r_1}})$ work for each node, although in practice 
 the order rarely needs to change.
 
\subsubsection{Propensity Bounds Given $u$}

 If we know $u$, the number of reaction events for $r_1$ and adjacent
 blocks, we can derive even tighter bounds on the reaction channels $R_{r_1 *}$. 
 In fact, these tighter bounds help us to efficiently increase $L$ when
 larger systems are considered, as will be demonstrated in section
 \ref{hierleap_derivation}.
 
 We determine $\bar{F}^{(r_1 *)}$ by finding bounds on species counts
 given $u$.
 In other words, we want to find
 \begin{equation*}
 n_A(r,n_0,k) \leq \bar{n}_{A r_1}(u, n_0) \text{, for }k=0..[(\text{index of
 final }r_1\text{ event}) - 1]
 \end{equation*}
which is the maximum possible value of $n_{A r_1}$ prior to the last
event in $r_1$ occurring. In this way $n_A(r,n_0,k)$ will never exceed the propensity
calculated from $\bar{n}_{A r_1}(u, n_0)$. 

Finding the optimal value for $\bar{n}_{A r_1}(u, n_0)$ is straightforward. 
We first need to consider blocks other than $r_1$ which 
may change $n_{A r_1}$. 
Since the order of reactions is unknown, we must assume
that all reaction events in blocks except $r_1$, written as
$u\backslash \{u_{r_1}\}$, occur prior to those in $u_{r_1}$.
It is desired that the number of neighbors relative to the total number
of blocks will be small. This will decrease $n_A(r,n_0,k)$ and ultimately 
lead to a more efficient algorithm. 
Secondly, we need to consider reactions in $r_1$. In calculating the bound, 
it is assumed that all $(u_{r_1}-1)$ reaction events chosen will behave
adversarially. This is analogous to the method considered in
section \ref{dtilde-hierleap}, with the modification that we will consider a subset of
reaction channels. Thus
\begin{equation*}
\bar{n}_{A r_1}(u,n_0) 
\equiv n_A + \displaystyle\sum_{{r_1}'}({u_{{r_1}'} -
\delta_{r_1 {r_1}'}}) \max_{{r_2}'} \Delta m^{({r_1}' {r_2}')}_{(a_1 a_2)}
\end{equation*}
will bound each $n_{A r_1}$ with respect to $r_1$ and $u$. 
The Kronecker delta function $\delta ( a,b)$ or $\delta ( a-b)$ is as usual:
\[
\delta ( a-b) =\delta _{a b}=\text{\boldmath $1$}\left( a=b\right)
\equiv \begin{cases}
1 & \mathrm{if}\  a=b \\
0 & \mathrm{otherwise} \\
\end{cases}.
\]
Finally, the propensities of reaction channels inside of block $r_1$ 
are bound as, 
\begin{equation*}
\bar{F}^{(r_1 r_2)}(u, n_0) \equiv F^{(r_1 r_2)}(u, \bar{n}_{A r_1}).
\end{equation*}

As in ER-leap, lower-bounding the propensities and species is done with the same techniques as that used for upper-bounding
with the restriction that propensities and species molecule counts cannot go below zero. 
These derived bounds are used in the following sections.

\subsection{Equivalent Markov Process}
\label{hierleap_derivation}

Similar to section \ref{erleap_markov_process}, we want to algebraically manipulate the distribution 
represented by the Chemical Master Equation (a special case of the Kolmogorov-Chapman equation \cite{ullah-2007-1})
into a form suitable for parallelization and acceleration. 
The hierarchical description from section \ref{hierarchical-notation} will aid
us in this transformation. 

First, note it is possible to rewrite equation (\ref{permutation_eq}) into a hierarchical version with $u$'s and $v$'s
strictly ordered such that
\begin{multline*}
\sum \limits_{\left\{ R_{k}|k=1 .. L-1\right\} }e( \mathbf{R}) =
\sum \limits_{\left\{ \mathbf{u} | u_{R}\in \mathbb{N}, \sum _{R}u_{R}=L\right\} }
\sum \limits_{\left\{ \mathbf{v_{r_1}} | v_{R}\in \mathbb{N}, \sum
_{r_2'}v_{r_1 r_2'}=u_{r_1}\right\} } \\
\sum \limits_{\left\{\sigma_1 | \sigma_1 \mathrm{\ permutes\ unequal\ }
R\mathrm{'s} | \mathbf{u}\right\} } 
\sum \limits_{\left\{\sigma_2  | \sigma_2 \mathrm{\ permutes\ unequal\ }
R\mathrm{'s} | \mathbf{v_{r_1}}\right\} } e( \sigma_1(\sigma_2(\mathbf{R})) ).
\end{multline*}
By taking an average of $e( \sigma(\mathbf{R}))$ and weighting
by the number of ways the selection may occur we get
\begin{multline*}
=\sum \limits_{\left\{ \mathbf{u} | u_{R}\in \mathbb{N}, \sum _{R}u_{R}=L\right\} }
\sum \limits_{\left\{ \mathbf{v_{r_1}} | v_{R}\in \mathbb{N}, \sum
_{r_2'}v_{r_1 r_2'}=u_{r_1}\right\} } 
{L \choose {R_1 R_2 \ldots R_n}}
\left< \left< e( \sigma_1(\sigma_2(\mathbf{R})) ) \right>_{\sigma_2} \right>_{\sigma_1} \\
\end{multline*}
and analogous to the way shuffling a deck of cards is the same as shuffling by
suit and then, maintaining that order, shuffling by value independently for each
suit, we may write
\begin{align*}
{L \choose {R_1 R_2 \ldots R_n}} &=
{L! \over {R_1! R_2! \ldots R_n!} } \\ 
&= {L! \over {u_{r_1}! u_{r_1'}! \ldots u_{r_1''}} } \prod_{r_1} 
{u_{r_1}! \over {v_{r_1 r_2}}!{v_{r_1 r_2'}}!\ldots{v_{r_1 r_2''}}!} \\
&={L \choose {u_{r_1} \ldots u_{r_1'}}} \prod_{r_1} 
{u_{r_1} \choose {v_{r_1 r_2} \ldots v_{r_1 r_2'}}} 
\end{align*}
and arrive at a useful form for our distribution, which is 
already suggestive of a block-parallel algorithm: 
\begin{multline}
\label{hier-leap-multiterms-avgs}
\sum \limits_{\left\{ r_{k}|k=1 .. L-1\right\} }e( \mathbf{r})
=\sum \limits_{\left\{ \mathbf{u} | u_{R}\in \mathbb{N}, \sum _{R}u_{R}=L\right\} }
{L \choose {u_{r_1} \ldots u_{r_1'}}}
\sum \limits_{\left\{ \mathbf{v_{r_1}} | v_{R}\in \mathbb{N}, \sum_{r_2'}v_{r_1 r_2'}=u_{r_1}\right\} }  \\
\prod_{r_1}
{u_{r_1} \choose {v_{r_1 r_2} \ldots v_{r_1 r_2'}}}
\left< \left< e( \sigma_1(\sigma_2(\mathbf{R})) ) \right>_{\sigma_2} \right>_{\sigma_1}
\end{multline}
To go further, we need to re-examine $e(\ldots)$.

\subsubsection{Introduction of Probability Bounds}

We now make use of our previously derived propensity bounds to 
derive a parallel algorithm. From equation (\ref{L_event_prob}) we have
\begin{equation*}
\left< \left< e( \sigma_1(\sigma_2(\mathbf{R})) ) \right>_{\sigma_2}
\right>_{\sigma_1} = \left< \left< \prod \limits_{k=L-1\searrow 0}
	\rho_{R_k}{F_{I_{k}(\mathbf{R},I_{0})}^{\left( R_{k} \right) }}
	\exp ( -\tau_{k}( D_{I_{k}( \mathbf{R},I_{0}) ,I_{k}( \mathbf{R},I_{0}) }
) )  \right>_{\sigma_2}
\right>_{\sigma_1};
\end{equation*}
with inclusion of derived bounds,
\begin{multline*}
\left< \left< e( \sigma_1(\sigma_2(\mathbf{R})) ) \right>_{\sigma_2}
\right>_{\sigma_1}
=\left< \left< 
\prod \limits_{k=L-1\searrow 0} \right. \right. 
{{F_{I_{k}(\mathbf{R},I_{0})}^{\left(R_{k} \right) }} \over 
			{\bar{F}^{(r_1 r_2)}(u, I_0)}}
{{\rho_{R_k}\bar{F}^{(r_1 r_2)}(u, I_0)} \over {\bar{D}^{(r_1)}(u, I_0)}} 
{{\bar{D}^{(r_1)}(u, I_0)} \over{\widehat{D}^{(r_1)}_{(I_0 L)}} }
\widehat{D}^{(r_1)}_{(I_0 L)}
  \times  \\
  \Bigg. \left.
	\exp ( -\tau_{k}( D_{I_{k}( \mathbf{R},I_{0})
			           ,I_{k}( \mathbf{R},I_{0}) } 
			           - \uhat{D}_{(I_0 L)}) 
    \exp(-\tau_{k}\uhat{D}_{(I_0 L)})
\Bigg>_{\sigma} \right>_{\sigma_1}.
\end{multline*}
If we separate out terms based on 
independence of $\sigma_1$, $\sigma_2$ and $v$, then
\begin{multline*}
\left< \left< e( \sigma_1(\sigma_2(\mathbf{R})) ) \right>_{\sigma_2}
\right>_{\sigma_1}
=
{\exp(-\mathbf{\tau}\uhat{D}_{(I_0 L)})}
\left( \prod_{r_1}{{}\widehat{D}^{(r_1)}_{(I_0 L)}}^{u_{r_1}}\right)
\left(
	\prod_{r_1} \prod_{r_2}
	{\left({{\rho_{R_k}\bar{F}^{(r_1 r_2)}(u, I_0)} \over 
	{\bar{D}^{(r_1)}(u, I_0)}} \right)}^{v_{r_2}} 
\right) \times \\
\left({\prod_{r_1}\left({{\bar{D}^{(r_1)}(u, I_0)} \over
{\widehat{D}^{(r_1)}_{(I_0L)}}}\right)^{u_{r_1}}} \right)
\times \\
\left< 
\prod \limits_{k=L-1\searrow 0} 
{{F_{I_{k}(\mathbf{R},I_{0})}^{\left(R_{k} \right) }} \over 
			{\bar{F}^{(r_1 r_2)}(u, I_0)}}
\left< 
	\exp ( -\tau_{k}( D_{I_{k}( \mathbf{R},I_{0})
          ,I_{k}( \mathbf{R},I_{0}) } 
          - \widehat{D}_{(I_0 L)}) 
\right>_{\sigma_2} \right>_{\sigma_1}\\
\end{multline*}
We now substitute the expression for 
$\left< \left< e( \sigma_1(\sigma_2(\mathbf{R})) ) \right>_{\sigma_2}\right>_{\sigma_1}$ 
into equation (\ref{hier-leap-multiterms-avgs}), combining terms where appropriate:
\begin{multline}
\label{hier-full-eq}
{\left[  \prod \limits_{k=l-1\searrow 0}\hat{W}\ \ \ \exp ( -\tau_{k}D)  \right] }_{I_{l}, I_{0}} = \\
  \left({\sum_{r_1'}\widehat{D}^{(r_1')}_{(I_0 L)} \over {\uhat{D}_{(I_0,L)}}}\right)^L
  \sum \limits_{\left\{ \mathbf{u} | u_{R}\in \mathbb{N}, \sum_{R}u_{R}=L\right\} } 
  \left[{L \choose {u_{r_1} \ldots u_{r_1'}}}
  			 \prod_{r_1} {\left({{{}\widehat{D}^{(r_1)}_{(I_0 L)}} \over \sum_{r_1'}\widehat{D}^{(r_1')}_{(I_0 L)}}\right)}^{u_{r_1}}
  \right]
\times \\
  \sum \limits_{\left\{ \mathbf{v_{r_1}} | v_{R}\in \mathbb{N}, \sum_{r_2'}v_{r_1 r_2'}=u_{r_1}\right\} }
  \left[
  \prod_{r_1}
  {u_{r_1} \choose {v_{r_1 r_2} \ldots v_{r_1 r_2'}}}
   \prod_{r_2}
	{\left({{\rho_{R_k}\bar{F}^{(r_1 r_2)}(u, I_0)} \over 
	{\bar{D}^{(r_1)}(u, I_0)}} \right)}^{v_{r_2}}  
  \right]
  \times \\
  {\left(\uhat{D}_{(I_0,L)}\right)}^L \exp(-\mathbf{\tau} \uhat{D}_{(I_0 L)})
  \times
  AcceptCoarse(u; I_0, L) \times \\
  \left( \prod_{r_1} AcceptBlock(v_{r_1}, \sigma_2; u) \right) \times
  AcceptFine(\sigma_1; u,v,I_0,\sigma_2)
\end{multline}
The acceptance probabilities are as follows:
\begin{equation}
  \label{accept-coarse}
  AcceptCoarse(u; I_0, L) = {\prod_{r_1}\left({{\bar{D}^{(r_1)}(u, I_0)} \over
{\widehat{D}^{(r_1)}_{(I_0L)}}}\right)^{u_{r_1}}} 
\end{equation}
\begin{equation}
\label{accept-block}
  AcceptBlock(v_{r_1}, \sigma_2; u, r_1) = 
  \prod_{k \in r_1}
     {{F_{I_{k}(\mathbf{R},I_{0})}^{\left(R_{k} \right) }} \over 
	 {\bar{F}^{(r_1 r_2)}(u, I_0)}}
\end{equation}
\begin{equation}
\label{pfine-accept}
AcceptFine(\sigma_1; u,v,I_0,\sigma_2) =
\prod \limits_{k=L-1\searrow 0} 
	\exp ( -\tau_{k}( D_{I_{k}
		( \mathbf{R},I_{0}) ,I_{k}( \mathbf{R},I_{0}) } - \widehat{D}_{(I_0 L)})) 
\end{equation}

Furthermore, prior to turning these equations into an algorithm, we note that we 
can lower-bound these acceptance probabilities. This will enable us to do an early 
acceptance or rejection without always doing all of the work to calculate these values exactly. 

\subsubsection{Lower Bounding Acceptance Probabilities}

We begin by lower-bounding $AcceptFine(\ldots)$. 
This probability requires the most work to calculate and as we will see may 
be bound fairly tightly. The bound only requires that $\mathbf{\tau}$ has been sampled.

The lower bound $\uhat{AcceptFine}(\ldots)$ is sought such that 
\begin{equation*}
\uhat{AcceptFine}(\ldots) \leq 
\prod \limits_{k=L-1\searrow 0} 
	\exp ( -\tau_{k}( D_{I_{k}
		( \mathbf{R},I_{0}) ,I_{k}( \mathbf{R},I_{0}) } - \uhat{D}_{(I_0 L)})).
\end{equation*}
for all possible $\{\tau_k\}$ and $\{I_0\ldots I_{L-1}\}$.
In the above, note that $\uhat{D}_{(I_0 L)}$ is constant with respect to $k$. 
Therefore, when we also upper bound 
\begin{equation*}
\widehat{D}_{I_0 L} \geq
D_{I_{k}( \mathbf{R},I_{0}) ,I_{k}( \mathbf{R},I_{0}) } ,
\end{equation*}
this creates an easily computable expression for the lower bound
\begin{equation*}
AcceptFine(\sigma_1; u,v,I_0,\sigma_2, \mathbf{\tau}) \geq
\prod \limits_{k=L-1\searrow 0} 
	\exp ( -\tau_{k}( \widehat{D}_{I_0 L} - \uhat{D}_{(I_0 L)})) 
\end{equation*}
so that 
\begin{equation}
\label{lower-bound-acceptfine}
\uhat{AcceptFine}(\tau; I_0,L) =
\exp ( -\mathbf{\tau}( \widehat{D}_{I_0 L} - \uhat{D}_{(I_0 L)}))	
\end{equation}
Furthermore, recall that $E[\mathbf{\tau}]={L / \uhat{D}_{(I_0,L)}}$. 
If we assume that $\Delta\widehat{D}_{(I_0 L)} \propto L$ when computing
$\uhat{D}_{(I_0 L)}$ and $\widehat{D}_{(I_0 L)}$ 
(see equation (\ref{near_optimal_block_bound})) in the limit of many non-zero propensity reaction channels
\begin{equation*}
\left< \lim_{|\mathcal{R}|\rightarrow\infty} \exp(-\mathbf{\tau}( \widehat{D}_{I_0 L} - \uhat{D}_{(I_0 L)}))\right>
\rightarrow 1
\end{equation*}
which implies that both $\uhat{AcceptFine}(\ldots)$ and $AcceptFine(\ldots)$  tend to unity as
the number of reaction channels increases.

Next, we set out to lower-bound $AcceptBlock(\ldots)$. This acceptance probability 
depends on $\sigma_2$. Therefore, work will be saved if we can calculate the lower bound
without sampling $\sigma_2$. This can be accomplished by noting that $AcceptBlock(\ldots)$ 
is a product of fractions. If we have a numerator and denominator that are independent of 
$\sigma_2$ we can re-write this equation in terms of $r_2$. Specifically, using 
$\ubar{F}^{(r_1 r_2)}_{u,I_0}$ allows us to lower-bound the equation.
\begin{equation*}
  AcceptBlock(v_{r_1}, \sigma_2; u, r_1) = 
  \prod_{k \in r_1}
     {{F_{I_{k}(\mathbf{R},I_{0})}^{\left(R_{k} \right) }} \over 
	 {\bar{F}^{(r_1 r_2)}(u, I_0)}}
\geq
  \prod_{k \in r_1}
     {{\ubar{F}^{(r_1 r_2)}(u,I_0)} \over 
	 {\bar{F}^{(r_1 r_2)}(u, I_0)}}
\end{equation*}
yielding 
\begin{equation}
\label{accept-block-low-bound}
\uhat{AcceptBlock}(v_{r_1}; u)=
  \prod_{r_2 \in r_1}
    {\left(
    	{{\ubar{F}^{(r_1 r_2)}{(u,I_0)}} \over 
	 	{\bar{F}^{(r_1 r_2)}(u, I_0)}}
	 \right)}^{v_{r_2}}
\end{equation}

\subsection{Algorithm}

The above equations, along with rejection sampling, allow us to 
create an efficient algorithm that will allow much of the work be
done in parallel. 
From equation (\ref{hier-full-eq}) observe there are two probability mass function expressions for a multinomial distribution.
Specifically, 
\begin{equation*}
Multinomial(u; 
\left\{ p_{r_1}={{{{}\widehat{D}^{(r_1)}_{(I_0 L)}} \over \sum_{r_1'}\widehat{D}^{(r_1')}_{(I_0 L)}}} \right\}, L)=
{L \choose {u_{r_1} \ldots u_{r_1'}}}
  			 \prod_{r_1} {\left({{{}\widehat{D}^{(r_1)}_{(I_0 L)}} \over \sum_{r_1'}\widehat{D}^{(r_1')}_{(I_0 L)}}\right)}^{u_{r_1}}
\end{equation*}
is the multinomial distribution for sampling $u$. And for each $r_1$ the vector $v_{r_1}$ is sampled as
\begin{multline*}
Multinomial(v_{r_1}; 
\left\{ 
   p_{r_1 r_2}=
	{{{\rho_{R_k}\bar{F}^{(r_1 r_2)}(u, I_0)} \over 
	{\bar{D}^{(r_1)}(u, I_0)}} } \right\}, u_{r_1}) = \\
 {u_{r_1} \choose {v_{r_1 r_2} \ldots v_{r_1 r_2'}}}
   \prod_{r_2}
	{\left({{\rho_{R_k}\bar{F}^{(r_1 r_2)}(u, I_0)} \over 
	{\bar{D}^{(r_1)}(u, I_0)}} \right)}^{v_{r_2}}
\end{multline*}
which is interesting and implies 
$v_{r_1}$ is independent of all other block's $v_{r_1'}$ given $u$. 

The multinomials $v_{r_1}$ may be sampled independently for each block, however it may be the case that equation (\ref{accept-block}) needs
to be computed by considering multiple blocks simultaneously. 
Specifically, computing $I_k(\mathbf{R}, I_0)$ for block $r_1$ may require knowledge about any neighboring blocks, $r_1'$, 
changing the chemical species counts for reaction channels $v_{r_1}$ reacting on the same species.
Since equation (\ref{accept-block-low-bound}) is independent of $I_k(\ldots)$, this ``joint'' acceptance probability 
only needs to be calculated when 
(a) blocks $r_1$ and $r_1'$ share a chemically reacting species in their respective $v_{r_1}$ and $v_{r_1'}$ 
sampled reaction channels and
(b) block $r_1$ or $r_1'$ does not pass early block-acceptance.
Computing equation (\ref{accept-block}) jointly involves computing a $\sigma_2$ for reaction events
in $r_1$ and $r_1'$. Then, $I_k(\ldots)$ may be computed properly.
It should be noted that in the pseudocode below some possible optimizations 
(eg independent early block accept and some parallel execution)
are not shown for clarity.
Instead ``connected components'', block groups which {\it may} need to be sampled jointly if conditions (a) and (b) are met, 
are sampled entirely in ``joint'' form.

We now present the HiER-leap algorithm, which is a realization of the aforementioned equations, in pseudocode.
First, note that if we have an early global acceptance, then most of the computational 
effort will be put into line 12 of the following pseudocode. 
The subroutine from this line will be shown later.
Notice that this function is independent for all blocks, with the exception that 
computing equation (\ref{accept-block}) {\it may} need to be done jointly for neighboring blocks,
 and needs to be done for all blocks with at least one reaction event.
This is an ideal scheme for parallelization and is done so with good efficacy as will be shown. 
Furthermore, for the tests in in section \ref{hier-acceleration} the full calculation $AcceptFine(\ldots)$ was rare because in general: 
$\uhat{AcceptFine}(\ldots) \geq 0.995$. The algorithm is as follows.

\begin{algorithmic}[5]

\Require $\widehat{D}_{(I_0,L)}, \uhat{D}_{(I_0,L)}, \{\widehat{D}^{(r_1)}_{(I_0,L)} \}$ precomputed for $L \geq 1$ {\textbf and} $I_0$.
\Ensure Return $(I_L, \Delta t)$ 
	\Comment return updated state and duration of $L$ steps

\Function{HiER-Leap}{}($I_0, L$)
\State $\mathbf{\tau} \gets$ \Call{Erlang}{}$(\mathbf{\tau}; \uhat{D}_{(I_0,L)}, L)$ 

\State $u \gets $\Call{Multinomial}{}$(u; 
\left\{ p_{r_1}={{{{}\widehat{D}^{(r_1)}_{(I_0 L)}} \over \sum_{r_1'}\widehat{D}^{(r_1')}_{(I_0 L)}}} \right\}, L)$

\State Compute $\bar{D}$'s \Comment May be done in parallel. 
\State \Comment In accordance with equation (\ref{accept-coarse}).
\If {\Call{UniformRandom}{0,1} $ \geq $\Call{AcceptCoarse}{}$(u; I_0, L)$} 
	\State \Return \Call{HiER-Leap}{}($I_0, L$) 
	\Comment Early Rejection. Try again.
\EndIf

\State 
\ForAll{$c \in $ \Call{ConnectedComponents}{}($u,R,I_0$)}
	\State \Comment See algorithm below.
	\State $(v_{r_1}, \sigma_2) \gets$ \Call{SampleConnectedComponents}{}($c, u, I_0$)
	\State \Comment{May be done in parallel.}
\EndFor

\State $z \gets $ \Call{UniformRandom}{0,1}
\If{$z \leq \uhat{AcceptFine}(\tau; I_0,L)$} 
		\Comment See equation (\ref{lower-bound-acceptfine}).
	\State \Return $(I_L(I_0,v,u), \tau)$
	\Comment Early Acceptance.
\EndIf

\State \Comment Computation should be rare; see equation (\ref{pfine-accept}).
\If {$z \leq$ \Call{AcceptFine}{}$(\sigma_1; u,v,I_0,\sigma_2, \mathbf{\tau})$} 
	\State \Return $(I_L(I_0,v,u), \tau)$
\Else
	\State \Return \Call{HiER-Leap}{}($I_0, L$) 
	\Comment Try again.
\EndIf

\EndFunction

\State
\Ensure Return each connected component of blocks, where two blocks are in the same connected 
component if they share a reaction species and each blocks has at least one reaction event 
such that  $u_{r_1}\geq 1$.
\Function{ConnectedComponents}{}($u,R,I_0$)
	\State $C \gets \{\}$
		\Comment $C$ is a set.
	\State $B \gets R$
		\Comment $B$ is the set of blocks.
	\While {$B \neq \varnothing$}
		\State $b \gets $ \Call{B.ChooseElement}{}() 
		\If {$u_b \geq 1$}
			\State $c \gets$ \Call{DepthFirstSearch}{}($b, u, B$)
				\State \Comment Blocks share an edge iff for each $u_{r_1}\geq 1$ and they share a species.
			\State $B \gets B\backslash c$ 
				\Comment Set operation subtraction.
			\State $C \gets $ \Call{C.Append}{c}
		\EndIf
	\EndWhile
	\State \Return $C$
\EndFunction 

\end{algorithmic}

The pseudocode to sample each connected component is as follows.

\begin{algorithmic}
\Require The connected component $c$ contains blocks which all have at least 
one reaction event.
\Function{SampleConnectedComponents}{}($c$, $u$, $I_0$)
	\State $pEarly \gets 1$
	\ForAll {$r_1 \in c$}
		\State $v_{r_1} \gets $\Call{Multinomial}{}
		$(v_{r_1}; 
			\left\{ 
   				p_{r_1 r_2}=
					{{{\rho_{R_k}\bar{F}^{(r_1 r_2)}(u, I_0)} \over 
					{\bar{D}^{(r_1)}(u, I_0)}} } 
			\right\}, 
			u_{r_1})$
		\For{$r_2 \in r_1$ \textbf{and} $v_{(r_1 r_2)} \geq 1$}
		\State $pEarly \gets pEarly \times    
						 		{\left(
    							{{\ubar{F}^{(r_1 r_2)}{(u,I_0)}} \over 
	 								{\bar{F}^{(r_1 r_2)}(u, I_0)}}
								\right)}^{v_{r_2}}$
		\EndFor
	\EndFor 
	 
	\State $z \gets $ \Call{UniformRandom}{0,1}
	\If {$z \leq pBlockEarly$}
		\State \Return $(\mathbf{v}, \sigma_2)$
		\Comment Early Accept.
	\EndIf  

	\State $pComponent \gets 1$
	\State $\sigma_2 \gets $ \Call{Permutation}{$u_{r_1}$}
	\Comment Must compute exact acceptance probability.
	\ForAll{$k = {1\ldots |\mathbf{v}|}$}
		\State $r_2' \gets \sigma_2(\mathbf{v}, k)$
		\State $pComponent \gets pComponent \times
     		{{F_{I_{k}({\sigma_2(v_{r_1})},I_{0})}^{\left(r_1 r_2' \right) } \over 
	 		{\bar{F}^{(r_1 r_2')}(u, I_0)}}}$
	 		\Comment Calculating $I_k$ takes the most work.
	\EndFor
	\If {$z \leq pBlock$}
		\State \Return $(\mathbf{v}, \sigma_2)$
		\Comment Accept sample.
	\Else
		\State \Return \Call{SampleConnectedComponents}{}($c$, $u$, $I_0$)
		\Comment Sample rejected, try again.
	\EndIf
\EndFunction
\end{algorithmic}

\section{Numerical Experiments}

\subsection{CaliBayes Validation}

We check the HiER-leap algorithm correctness numerically with the CaliBayes test suite similar to the work in ER-leap
\cite{Mjolsness_Orendorff_Chatelain_Koumoutsakos_2009}.
If is possible to solve
analytically for $P(X|t)$, this allows us to compare many simulated trajectories to the true distribution defined by the CME.
Since HiER-leap reduces to ER-leap when the number of blocks goes to one, and it has already been show that ER-leap samples the correct distribution, 
we wish to test across a variety of reaction channel quantities and organization structure. 

The reaction networks in CaliBayes for which we know the analytical solution involve at most two species types. 
However, simulating many replicates of these networks on a grid, not connected with diffusion, 
will allow us to treat each block as an independent sample. We can then treat the simulation of many network replicates 
as many sampled trajectories of a single network. 

We perform tests over a number of network replicates $m=2\ldots1000$. The number of blocks range from $b=1\ldots m$. 
The leap is in the range $L=3\ldots18$, where the leap used depends on the specific reaction network, $m$ 
and $b$, but is held constant throughout the simulation.

CaliBayes models 1-01, 1-03, 1-04, 2-01, 2-02, 3-01 and 3-02 \cite{CaliBayes2008}
are tested, on the spaced defined by the Cartesian product of the possible values for the $m$, $b$ and $L$ parameters as described above,
for parameters which result in an acceptance probability greater than about $0.05$. 
These tests pass on these cases using the criteria of Evans et al. \cite{CaliBayes2008}

We now turn to a large, spatially coupled system.

\subsection{Acceleration}
\label{hier-acceleration}

\begin{figure}[h]
\begin{center} 
\includegraphics[width=3.375in]{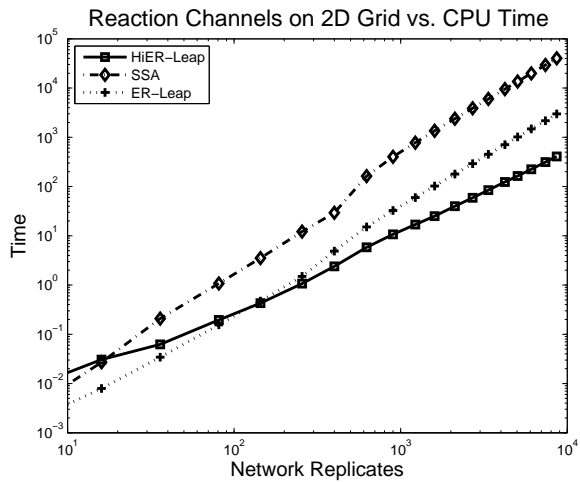}
\end{center}
\caption{The Williamowski-R\"ossler model as seen in 
section \ref{hier-acceleration} is used for this experiment. There are different number of network replicates on a 2D square grid with 
diffusion rate of 0.1. The number of replicates ranges from $4$ to $8649$ which equates to $64$ to $189612$ reaction channels.}
\label{hierChannelsScale}
\end{figure}

\begin{figure}[h]
\begin{center} 
\includegraphics[width=3.375in]{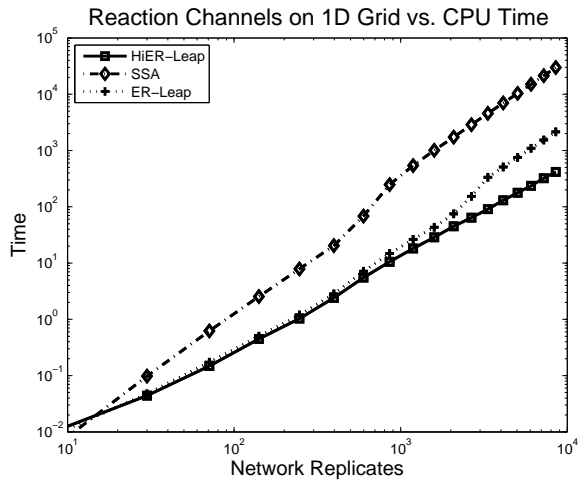}
\end{center} 
\caption{The same experimental setup as 
used for figure \ref{hierChannelsScale} except 1D diffusion is used.} 
\label{hierChannelsScale1D}
\end{figure}

As an exact algorithm, the key performance metric of relevance to HiER-leap is the amount of acceleration achievable. 
As discussed earlier, in principle adding more reaction channels and processors should increase the relative speedup over SSA. 
We can see this trend experimentally in figure \ref{hierChannelsScale} and figure \ref{hierChannelsScale1D}.

\newpage

We test using a spatially coupled version of the Williamowski-R\"ossler model \cite{WangLi98} defined as 
\begin{gather*}
 X\overset{k_{1}}{\operatorname*{\longleftrightarrow }\limits_{k_{2}}}2X
\quad Y\overset{k_{5}}{\operatorname*{\longleftrightarrow }\limits_{k_{6}}}\emptyset
\quad Z\overset{k_{9}}{\operatorname*{\longleftrightarrow }\limits_{k_{10}}}2Z
\\
 X+Y\overset{k_{3}}{\operatorname*{\longleftrightarrow }\limits_{k_{4}}}2Y
\quad X+Z\overset{k_{7}}{\operatorname*{\longleftrightarrow }\limits_{k_{8}}}\emptyset
\end{gather*}
replicated over a $d$-dimensional grid for $d=1$ or $d=2$. 
Diffusion reaction channels for all species are added between adjacent grid cells with a rate of $k_d=0.1$.  
Parameters and initial conditions for each of the replicated Williamowski-R\"ossler grid cells are as follows: $k_{1}=900$, $k_{2}=8.3 \times
10^{-4}$, $k_{3}=0.00166$, $k_{4}=3.32 \times 10^{-7}$, $k_{5}=100$, $k_{6}=18.06$,
$k_{7}=0.00166$, $k_{8}=18.06$, $k_{9}=198$, $k_{10}= 0.00166$.
$X(0)=39570$. $Y(0)=511470$. $Z(0)=0$.

The following tests are all run on an Apple Macintosh Pro with a Quad-Core Intel Xeon processes running a total of 8 cores at 2.26 GHz and 13 GB of
RAM using OS X 10.6.8. The algorithms are coded in C++ and $Boost.Thread$ \cite{Kempf:2002:BTL} and the Intel Threading Building Blocks 
\cite{intel-tbb} are used for
multithreading. 
Connected components were found using the depth-first search algorithm.
We compiled the code using the LLVM compiler 1.0.2. The HiER-leap code may be found at  
 {\it http://computableplant.ics.uci.edu/hierleap/}. 
 
 Results are shown in figures \ref{hierChannelsScale} and \ref{hierChannelsScale1D}.
They show a substantial speedup of HiER-leap over SSA and ER-leap, around 100x and 10x respectively, as we increase the number of
 reaction channels to around 190,000 . 
 The spatial nature of this experiment means that blocks are neighbors with relatively few other blocks. 
 This leads to a greater ``coarse-scale'' acceptance probability and therefore increased efficiency. 

Additionally, we see that the slopes of the log-log runtime plots for SSA and ER-leap become nearly
equal as the number of reaction channels increase. 
This is expected, since ER-leap finds bounds on individual reaction channels after $L$ reaction events, 
and this bound is independent of the number of reaction channels. 
HiER-leap however does not have this shortcoming and has a lower slope (eg better asymptotic behavior) as a result.

\subsection{HiER-leap Properties}

The algorithm parameters, such as leap size and hierarchical organization, require optimization 
before the fastest possible execution time is achieved.
To find the ideal methods with which to optimize our algorithm, we explore various trade-offs here.

\begin{figure}[h]
\begin{center} 
\includegraphics[width=3.375in]{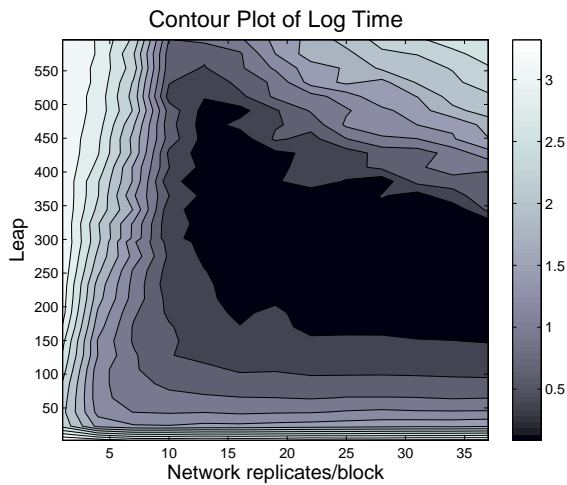}
\end{center} 
\caption{Log CPU Time vs Leap and Hierarchical Structure. The Williamowski-R\"ossler model as seen in 
section \ref{hier-acceleration} is used for this experiment. There are 400 network replicates on a 1D grid with 
diffusion rate of 0.1. 
The model execution time depends on leap and hierarchical organization. 
As leap increases the amount of work per iteration goes up but the acceptance ratio goes down.
Furthermore, if there are many reaction channels per block the total acceptance probability of the system goes 
down. However, in this situation the inner-block acceptance probability goes up. 
When the number of reaction channels per block goes down, the opposite trends occur. 
In this way the chosen leap and block organization will determine the total execution time.}
\label{HierLogContures}
\end{figure}

In figure \ref{HierLogContures} we observe that the optimal $b$ and $L$ are interdependent
for a given network. 
However, it is interesting to note that for this experiment there is a relatively large plateau 
of nearly equivalent optimal running times. This means that the range of reasonably good 
parameters is large. 
Furthermore, the contour plot of figure \ref{HierLogContures} indicates that there is only one global 
optimum. This seemingly convex behavior indicates that finding the optimum requires
only a simple hill climbing algorithm. 

Thus, the results from figure  \ref{HierLogContures} indicate that finding the optimal $L$ and hierarchical 
organization for a spatially distributed system is an easy optimization problem. 
These results, and those from ER-leap, suggest that $L$ will generally have a local optimum that is also a global optimum. 
However, the optimal configuration of the blocks and reaction channels for networks not specifically representing a 
spatially distributed reaction network
remains an open problem.

\section{Summary}

We have presented a novel accelerated stochastic simulation
algorithm which has demonstrated an ability to sample from the CME
without a loss of accuracy. 
Due to its hierarchical design, 
this method 
(a) scales very well with the number of reaction channels and simultaneously 
(b) takes 
advantage of parallel hardware for single trajectory samples. 
As far as we are aware, this is the first exact accelerated algorithm 
with either property (a) or (b),
and is therefore of potential significance to the computational biology community. 

Open questions and future work abound. 
For example, it is not know how well this method works on `real networks' of substantial complexity
taken from biological modeling practice. 
We believe that modular structure in biological networks
will make the method particularly useful. 
Additionally, it is unknown how substantial increases in the 
parallel architectures of future computers
will increase performance. 
 
\section*{Acknowledgements}

We acknowledge useful discussions with Petros Koumoutsakos.
Funding was provided by US NIH P50-GM76516, R01-GM086883,
and US NSF \#EF-0330786.
 
\appendix

\section*{Appendix}
\label{max-delta-proof}
We will show that for ${\Delta \widehat{D}^{(r_1)}_{(I_0 L)}}^*$ from equation (\ref{near_optimal_block_bound})
and $\Delta \widehat{D}^{(r_1)}_{(I_0 L)}$ from equations (\ref{delta_used_for_alg}) and
(\ref{delta_used_for_alg2}),
it is the case that ${\Delta \widehat{D}^{(r_1)}_{(I_0 L)}}^*\leq \Delta \widehat{D}^{(r_1)}_{(I_0 L)}$. 
  
\begin{proof}[Proof by contradiction.]

Assume there is some $r_2\in r_1$ and state $I'=\mathbf{n}'$ with
$\forall_a n_a' \leq \tilde{n}_a$ and $\exists_a n_a' < \tilde{n}_a$, 
reachable from $I_0$
in at most $L-1$ reaction events used to find ${\Delta \widehat{D}^{(r_1)}_{(I_0 L)}}^*$ such that
${\Delta \widehat{D}^{(r_1)}_{(I_0 L)}}^*> \Delta \widehat{D}^{(r_1)}_{(I_0 L)}$. 
Substituting in our definitions for ${\Delta \widehat{D}^{(r_1)}_{(I_0 L)}}^*$ and $\Delta \widehat{D}^{(r_1)}_{(I_0 L)}$ , 
using equation \ref{D_compute_eq}, and introducing the notation that $I(r_2)$ will be the result of $r_2$ applied to $I$ and $I({r_2}^+)$ 
is the result of  $r_2$ applied to $I$ only for species which have net gain ($\Delta m^{(r_1r_2)}_a>0$),
yields
\begin{align*}
{\Delta \widehat{D}^{(r_1)}_{(I_0 L)}}^*&=D^{(r_1)}_{I'(r_2)}-D^{(r_1)}_{I'} \\
&=\sum_{r_2'' \in r_1} \rho_{(r_1r_2'')} F^{(r_1r_2'')}_{I'(r_2)} - \sum_{r_2'' \in r_1} \rho_{(r_1r_2'')} F^{(r_1r_2'')}_{I'} \\
&=\sum_{r_2'' \in r_1} \rho_{(r_1r_2'')} \left(  F^{(r_1r_2'')}_{I'(r_2)} -  F^{(r_1r_2'')}_{I'} 
\right)
\end{align*}
and
\begin{equation*}
{\Delta \widehat{D}^{(r_1)}_{(I_0 L)}}=\sum_{r_2'' \in r_1} \rho_{(r_1r_2'')} \left(  
F^{(r_1r_2'')}_{\tilde{I}({r_2}^+)} -  F^{(r_1r_2'')}_{\tilde{I}} 
\right).
\end{equation*}
Therefore, we can equivalently say that we are trying to disprove 
\begin{multline}
\label{single_reaction_disprove}
\sum_{r_2'' \in r_1} \rho_{(r_1r_2'')} 
   \left(  F^{(r_1r_2'')}(\mathbf{n}' + \mathbf{\Delta m}^{(r_1 r_2)}) -  F^{(r_1 r_2'')}(\mathbf{n}')
   \right) > \\
\sum_{r_2'' \in r_1} \rho_{(r_1r_2'')} 
   	\left(  
		F^{(r_1 r_2'')}(\tilde{\mathbf{n}} +\mathbf{\Delta m}^{(r_1r_2)}) -
        F^{(r_1 r_2'')}(\tilde{\mathbf{n}})
	\right).
\end{multline}
Note that by grouping terms by $r_2''$, there is a one-to-one correspondence between the summation terms on each side of the 
inequality. 
   
If true, equation (\ref{single_reaction_disprove}) implies that there is at least one reaction channel $r_2' \in r_1$ for $\Delta m^{(r_1r_2)}_a>0$
such that
\begin{equation}
\label{difference_inequality}
F^{(r_1 r_2')}(\mathbf{n}' + \mathbf{\Delta m}^{(r_1r_2)}) - F^{(r_1 r_2')}(\mathbf{n}') > 
F^{(r_1 r_2')}(\tilde{\mathbf{n}} +\mathbf{\Delta m}^{(r_1r_2)}) -
F^{(r_1 r_2')}(\tilde{\mathbf{n}}).
\end{equation}
But we will show that this is impossible for any $\tilde{n}_a > n_a'\geq 0$. 
Note that we do not need to consider $\Delta m^{(r_1r_2)}_a\leq0$ because $F$ is monotonic, the LHS will be decreased and 
the RHS will not change as per the definition of $\Delta \widehat{D}^{(r_1)}_{(I_0 L)}$ (negative $\Delta m^{(r_1r_2)}_a$ are ignored). 

Before proceeding we will introduce the forward difference operator, $\Delta_{F(i)}$, 
such that
\begin{align} \Delta_{F(i)} f(z) &\equiv f(z+i)-f(z) \label{difference_operator} \end{align}
for any function $f(z)$. 

Furthermore, $F^{(r_1r_2)}(\mathbf{n})$ can be decomposed by species into terms including chemical species $C_a$ and those 
which do not. Following from equation (\ref{F_eq}), this allows us to rewrite $F^{(r_1r_2)}(\mathbf{n})$
 as
\begin{equation*}
\label{const_term}
F^{(r_1 r_2')}(\mathbf{n})=
G^{(r_1r_2')}(\mathbf{n}\backslash\{n_a\}) \times (n_a)_{k}
\end{equation*} 
for some constant $G^{(r_1r_2)}(\mathbf{n}\backslash\{n_a\}) \geq 0$ which does not depend on $n_a$, 
where $k=m^{(r_1r_2')}_a$ is the input stoichiometry for reaction
$r_2'$ and species $C_a$, and
\begin{equation*}
(n)_k \equiv \frac{n!}{(n-k)!}. 
\end{equation*} 


For equation (\ref{difference_inequality}) to be true there must exist a species $C_a$ such that 
\begin{equation}
\label{single_species_delta_toprove}
F^{(r_1 r_2')}(n_a' + \Delta m^{(r_1 r_2)}_a) - F^{(r_1 r_2')}(n'_a) > F^{(r_1 r_2')}(\tilde{n}_a +\Delta m^{(r_1 r_2)}_a) - 
F^{(r_1 r_2')}(\tilde{n}_a)
\end{equation}
is true. All of the above $F^{(r_1r_2)}$ are calculated using ${n_b=\mathbf{n}'\backslash\{n_a\}}$ and $n_a\in \{n_a', \tilde{n}_a\}$.
When we show that $n_a'$ will not result in a greater delta than that offered by using $\tilde{n}_a$ instead, 
this implies that equation (\ref{difference_inequality}) may never be true.

Equivalent to equation (\ref{single_species_delta_toprove}), by dividing out $G^{(r_1r_2)}(\mathbf{n}') \geq 0$, using equation
(\ref{difference_operator}), and setting $m=\Delta m^{(r_1 r_2)}_a$ we arrive at
\begin{equation}
\label{show_false}
\Delta_{F(m)}(\tilde{n}_a)_k - \Delta_{F(m)}(n_a')_k < 0.
\end{equation}

However, because $n_a'<\tilde{n}_a$, if it is shown that $\Delta_{F(m)}(n)_k$ is monotonic in $n$ then this will imply
equation \ref{show_false} is false.

Therefore, it just remains to be shown that $\Delta_{F(m)}(n)_k$ is monotonic in $n$. 
Consider the following equation which tests for monotonicity
\begin{align*}
\Delta_{F(m)}&(n+1)_k-\Delta_{F(m)}(n)_k \\
&=\left[\Delta_{F(1)}(n+m)_k+\ldots+\Delta_{F(1)}(n+1)_k\right]-\\
&\ \ \ \  \ \ \ \left[ \Delta_{F(1)}(n+m-1)_k+\ldots+\Delta_{F(1)}(n)_k
\right] \\
&=\Delta_{F(1)}(n+m)_k-\Delta_{F(1)}(n)_k \\
&=k(n+m)_{k-1}-k(n)_{k-1} \\
&=k \left[ \frac{(n+m)!}{(n+m-k+1)!} - \frac{n!}{(n-k+1)!} \right] \\
&=k\frac{n!}{(n-k+1)!} \left[ 
\frac{n+m}{n+m-k+1}\times \ldots \times \frac{n+1}{n-k+2}-1
\right] \\
&\geq 0
\end{align*}
because $k \geq 1$ implies every factor in
the long product is $\geq 1$. This implies monotonicity.
Therefore equation (\ref{single_species_delta_toprove}) is false for all $C_a$, implying equations (\ref{difference_inequality}) is false,
as was to be proved.

\end{proof}   

\bibliographystyle{abbrv} 
\bibliography{hier}

\end{document}